\documentclass[12pt,reqno]{amsart}
\usepackage{amsmath,amsfonts,amsthm,amssymb}
\usepackage{amsxtra,graphicx,hyperref}

\renewcommand{\baselinestretch}{1.1}
\newtheorem{thm}{Theorem}
\newtheorem{cor}{Corollary}
\newtheorem{lem}{Lemma}
\newtheorem{prop}{Proposition}
\theoremstyle{definition}
\newtheorem{defn}{Definition}
\newtheorem{rem}{Remark}

\newcommand{\sima}{\mathrel{\rlap{\hbox{$\sim$}}\raise.95ex\hbox{{\tiny
 \,A}}}}

\newcommand{\simt}{\mathrel{\rlap{\hbox{$\sim$}}\raise.95ex\hbox{{\tiny
\,\,\textrm{T}}}}}

\newcommand{\beq}{\begin{equation}}
\newcommand{\eeq}{\end{equation}}

\usepackage{graphicx, epstopdf}

\usepackage{caption}
\usepackage{subcaption}

\usepackage{latexsym}

\begin{document}

\title{A Direct Road to Entropy and the Second Law of Thermodynamics}

\author{Jakob Yngvason}
\address{Jakob Yngvason, Faculty of Physics, University of Vienna, Boltzmanngasse 5, 1090 Vienna, Austria}
\email{jakob.yngvason@univie.ac.at}

\begin{abstract} The entropy of classical thermodynamics is uniquely determined by the relation of adiabatical accessibilty between equilibrium states of thermodynamical systems. This review outlines the logical path leading to this results and the challenges that have to be faced on the way.

\end{abstract}



\maketitle

\section{Introduction}
This contribution to the {\em Festschrift} for Elliott Lieb is conceived as a concise review of the axiomatic approach to entropy and the second law of classical thermodynamics developed by Elliott and the present authors in \cite{LY99} and elaborated in subsequent papers \cite{LY00a}-\cite{LY03}. These  papers  occupy a special place in Elliott's {\oe}vre since their conceptual and mathematical framework is rather different from that of his other works, most of which concern  (many-body) quantum theory and statistical physics. 

The aim of the present expos\'e is to explain in a succinct way the logical structure of this analysis. 
It is hoped that it may be useful as a summary e.g. for students of thermodynamics and raise their appetite for taking a look at  the original sources where detailed explanations of the basic concepts are presented together with full mathematical proofs of all statements. Reviews that are intermediate in length between the present short summary of the theory and the full version can be found in in \cite{LY98}, \cite{LY00a}, \cite{LY01} and \cite{LY03};  the latter two include  complete proofs of most of the theorems mentioned in the sequel. The Introduction to \cite{LY99} has also a discussion of the relation of our work to previous work in a similar spirit, in particular 
 the great book of Robin Giles \cite{Giles},  and also of other views on the foundations of thermodynamics.  References to some recent works with alternative approaches are mentioned in the last Section.  

The starting point of our investigations is the empirical fact that under {\em adiabatic} conditions-- a concept that will be made more  precise below-- certain changes of the equilibrium states of thermodynamical systems are possible and some are not.
A basic task of thermodynamics -- as we see it-- is to {\em separate the possible from the impossible in a quantitative way}.  From this point of view the second law of thermodynamics is encapsulated in the \textbf {Entropy Principle:} \medskip

{\em For equilibrium states of macroscopic systems the possible adiabatic state changes are characterized by the increase (non-decrease) of an (essentially) unique state function, called {\textbf entropy}, that is extensive and additive on subsystems. }
\medskip

I emphasize right away that the word \lq\lq adiabatic\rq\rq\  is here not used in sense of \lq\lq slow\rq\rq\ as it often means in other disciplines of physics. Our operational definition of the concept is as follows: 

\textit{ A change of a state $X$ to a state $Y$ of a thermodynamic system is called adiabatic if its only {\em net} effect on the surroundings of the system is that a weight may have risen or fallen in a gravitational field.}\medskip

In this case we say that $Y$ is \textbf{adiabatically acccesible} from $X$ and write this symbolically as
\beq X\prec Y\eeq
(read \lq $X$ precedes $Y$\rq.) 

The phrase \lq\lq under adiabatic condition\rq\rq\  used above means that the state change is adiabatic in the sense of the operational definition. Note that in this definition there is no mention  of \lq\lq thermal isolation\rq\rq\  or \lq\lq heat\rq\rq. In fact, in contrast to traditional approaches, these are not basic concepts in our framework.

The consequences of the entropy principle reach far beyond the historical roots of thermo\-dynamics as an analysis of the efficiency of thermal engines. It is \textit{all important for understanding the properties and behaviour of macroscopic material systems}. 
In this context I take the opportunity to recommend the excellent textbook by  Andr\'e Thess \cite{AT07}, which contains an abbreviated version of our framework together with many instructive applications in physics and engineering. In fact, our approach has found its way into several engineering curricula at German universities, in particular Ilmenau University of Technology and Stuttgart University.

\medskip

Our road to entropy and the second law can be outlined as follows:

In the next section we start by putting forward six general axioms, denoted A1-A6, for the relation $\prec$. These axioms are very plausible, even self evident,  bearing in mind the physical interpretation of the relation. We then prove in Theorem 1 below that  A1-A6, if assumed {\em together} with a property we call the \textit {comparison hypothesis} (CH), are \textit {equivalent} to the entropy principle. (In brief, the comparison hypothesis requires that for any pair of states $X$, $Y$ under consideration at least one of the alternatives $X\prec Y$ or $Y\prec X$ holds.) This part of the analysis concerns a general mathematical question about order structures and results similar to our Theorem have, in fact, appeared both earlier and later in quite different contexts, e.g., in financial mathematics  \cite{vN}, \cite{HM}, \cite{D}. Such structures have also a bearing on modern \textit{resource theories} \cite{Ch} and certain aspects of \textit{quantum information theory}, see, e.g.,  \cite{Weile} and references cited therein.

Theorem 1 exhibits clearly which mathematical properties of the relation $\prec$ are needed to derive the entropy principle. It  leads directly to an explicit  formula for entropy, Eqs. \eqref{1.17} and \eqref{1.18} below,  {\em expressed solely in terms of the relation $\prec$}.  
This formula, remarkable as it is, does not end our story, however. For one thing, we do not consider it natural to accept CH as an axiom without proof and a substantial part of our further analysis consists in deriving it from physically more transparent assumptions. These involve familiar concepts from traditional thermodynamics, in particular energy, volume and pressure, and last but not least, that of thermal equilibrium, which can be defined using the relation $\prec$.   Besides proving the comparison hypothesis, a mathematically rigorous derivation of the standard practical recipes for measuring and computing entropy and absolute temperature in terms of these  concepts is an essential part of our analysis of the foundations of thermodynamics.

In  Section 3 we single out a class of thermodynamical systems which we call  \textit{ simple systems}. Here coordinates appear for the first time so that the state space of a simple system can be regarded as a subset of $\mathbb R^{n+1}$ for some $n\geq 1$. One coordinate, the  internal {\em energy}, denoted by $U$, plays a distinguished role, the other $n$ coordinates are called {\em work coordinates} (another common term is {\em deformation coordinates}), the volume being a primary examples. The possibility of using the energy $U$ as a coordinate is justified by reference to the  first law of thermodynamics. 

 An assumption, which is physically motived and simplifies greatly the mathematical analysis, is the requirement that formation of convex combinations, in the sense of the usual convex structure of $\mathbb R^{n+1}$, is an adiabatic operation in the sense of the relation $\prec$.  We denote this assumption by A7.
 
The  axioms A1-A7 are so general that they cover also the trivial case that $X\prec Y$ for all states $X$ and $Y$. The entropy would then simply be a constant and the entropy principle vacuous.  We  now assume an axiom for simple systems, denoted S1, demanding  that for every $X$ there is a $Y$ such that $X\prec Y$, but $Y\prec X$ is {\em not} true. Taken together with the convexity assumption this turns out to be equivalent to {\em Caratheodory's principle} \cite{Cara09}, namely in every neighbourhood of every state there are states that are \textit{not} adiabatically accessible from it. This establishes a direct connection with traditional formulations of thermodynamics and together with some further natural assumptions, denoted S2-S3, leads to the conclusion that  two states $X$, $Y$ of the same simple system are always comparable, i.e., one of the alternatives $X\prec Y$ or $Y\prec X$ must hold.

The comparison hypothesis CH requires, however,  more than the comparability  of states of a single simple system.  The definition itself and the important property of {\em additivity} of entropy involves comparison of  states of \textit{compounds } of simple systems. An essential new ingredient is the possibility to connect  two simple systems by a  {\em thermal join} to form a new simple system. This leads naturally to new axioms about thermal contact, denoted T1-T5, and the concept of thermal equilibrium. Axiom T3 in particular is the {\em zeroth law} of thermodynamics. With the aid of the axioms A1-A7, S1-S3 and T1-T5 we can derive the entropy principle for all adiabatic state changes which do not involve mixing of substances or chemical reactions. The entropy is uniquely determined, up to one multiplicative constant and one additive constant for each basic simple system with a fixed chemical composition. Last but not least, our axioms imply that entropy is a continuously differentiable function of the energy and the work coordinates, and that  absolute temperature, characterizing thermal equilibria,  can be defined as the reciprocal of the derivative of entropy with respect to the energy. 

A final chapter of our analysis concerns state changes involving mixing of different substances as well as chemical reactions. Traditional treatments rely here on the existence of {\em semipermeable membranes} which, however, are idealizations without real counterparts except in a few special cases. Nevertheless without invoking fictious semipermeable membranes (but adding a final axiom about nonexistence of \lq sinks\rq\  from which matter cannot escape) we are able to derive quite generally  a slightly weakened form of  entropy principle  and identify experimentally testable conditions under which the full principle holds.

\section{The general axioms and the entropy formula}

The basic concepts entering the general axioms are thermodynamic \textit{states} and \textit{state spaces} together with two elementary operations: \textit{Composition} and \textit{scaling}. 

Mathematically, state spaces are just sets,  $\Gamma_1$, $\Gamma_2$...   with states, $X$, $Y$, $Z$,... as their elements. The composition (or product) of two
state spaces $\Gamma_1$ and $\Gamma_2$ is a new \textit{compound state space}, defined  as the cartesian product $\Gamma_1\times \Gamma_2$ of all pairs $(X,Y)$ with $X\in\Gamma_1$, $Y\in \Gamma_2$.
 One can also compose more that two state spaces, $\Gamma_1,\Gamma_2, \Gamma_3,...$, to form $\Gamma_1\times \Gamma_2\times\Gamma_3\times\cdots$.  

A  \textit{scaled copy} of a state space $\Gamma$ for $\lambda>0$ is a new state space denoted $\lambda\Gamma$ \footnote{in \cite{LY99} the notion $\Gamma^{(\lambda)}$ is used} with states denoted $\lambda X$, $X\in\Gamma$.  State spaces of the type $\lambda_1\Gamma\times \lambda_2\Gamma\cdots $
with $\lambda_i>0$ are called \textit{multilple scaled copies} of $\Gamma$. We can trivially allow also $\lambda_i=0$ for some $i$ by simply dropping the corresponding $\lambda_i\Gamma$ from the cartesian product. We shall refer to the number $\lambda_1+\lambda_2+\cdots$ as the total \textit{matter content} of the states in the multiple scaled copy of $\Gamma$. A function $F$ defined on multiple scaled copies of $\Gamma$
is called \textit{additive}  if \beq F(X,Y)=F(X)+F(Y)\eeq and \textit{extensive} if
\beq F(\lambda X)=\lambda F(X).\eeq An 
\textit{additive} \textit{and} \textit{extensive}  state function thus satisfies
\beq F(\lambda_1X_1, \lambda_2X_2,\dots)=\sum _i\lambda_i F(X_i)\eeq
for all $X_i\in\Gamma$ and $\lambda_i\geq 0$.

So far, states and state spaces are just abstract symbols satisfying some self-evident rules of algebraic manipulations that we do not list here. (See pp. 14 and 15  in  \cite{LY99}.) Physically, composition means that we put two systems side by side on a laboratory desk and regard them as a new system. Scaling means that the matter content is multiplied by the parameter $\lambda$ and likewise extensive state variables like energy and volume. The mathematical theorem to follow does not depend on this interpretation, however.

The theorem concerns a relation $\prec$ defined for pairs of states. We shall use the following nomenclature: If $X\prec Y$ we say that $Y$ is {\em adiabatically accessible} from $X$, or that $X$ \textit{precedes} $Y$. 
We say that two states, {$X$} and {$Y$} (not necessarily in the same state space)  are \textit{comparable} if \beq
\hbox{either {$X\prec Y$} or {$Y\prec X$} (or both).}\eeq
 The states are \textit{adiabatically equivalent}, written \beq X\sima Y,\eeq if both conditions hold. 
  If {$X\prec Y$} but {$Y\not \prec X$} we say that {$X$} 'strongly precedes' {$Y$} and write {\beq X\prec\prec Y.\eeq}


The general assumptions about the relaton $\prec$ are as follows:
\begin{itemize}
\item[\textbf {A1.}]  \textit{ Reflexivity\/}:  {$X \prec X$}
\item[{\textbf A2.}]  \textit{ Transitivity:\/} If {$X \prec Y$} and {$Y \prec Z$}, then {$X 
\prec Z$}.

\item[{\textbf A3.}] \textit{ Consistency\/}: If {$X \prec X^\prime$} and {$Y 
\prec Y^\prime$}, then {$(X,Y) \prec
(X^\prime, Y^\prime)$}

   \item [{\textbf A4.}] \textit{ Scaling Invariance\/}: If {$\lambda > 0$} and
{$X,Y \in \Gamma$} with {$X \prec Y$},  then
{$\lambda X \prec \lambda Y$}

\item[{\textbf A5.}]  \textit{ Splitting and Recombination\/}:
{$X \prec ((1-\lambda) X, \lambda X)\prec X.$}

\item[{\textbf A6.}]  \textit{ Stability\/}: If
{$(X, {\varepsilon} Z_0) \prec (Y, {\varepsilon} Z_1)$}
for some {$Z_0$}, {$Z_1$} and a sequence of {$\varepsilon$}'s tending to zero, then
{$X \prec Y$}. 
\end{itemize}


Conditions (A1)-(A6) are all 
highly plausible if {$\prec$} is interpreted as the relation of adiabatic 
accessibility in the sense of the operational definition given in the Introduction.
They are also clearly {necessary},  but still {not sufficient} for the existence of 
an additive and extensive entropy that characterizes the relation on compound systems made of 
scaled copies of {$\Gamma$}. A further property is needed in order to arrive at an entropy:
\begin{itemize}
\item [{$\textbf {CH}$.}] \textit{Comparison Hypothesis} for a state space $\Gamma$:
Any two states in the collection of state spaces {$(1-\lambda)\Gamma\times \lambda \Gamma$} with $0\leq \lambda\leq 1$ are comparable.
\end{itemize}
\rem The {$\mathrm  {CH}$} implies that all states in a multiple scaled copy of $\Gamma$ are comparable, provided they have the same matter content as defined above. This is a simple consequence of assumption A1-A5. This is a stronger requirement than comparabilty for states in $\Gamma$ alone as discussed in Remark 3 below.


\begin{thm}[\textbf{Existence and uniqueness of entropy, given CH}] The following properties are  \textit{equivalent} for a state space $\Gamma$:
\begin{itemize}
\item[(1)] The relation {$\prec$} satisfies assumptions A1-A6 and $CH$.
\item[(2)] There is a function, called \textit{entropy} and denoted by $S$, defined on all multiple scaled copies of\/ $\Gamma$ such that, whenever two states $X$ and $Y$ have the same  matter content,
\beq X\prec Y \quad \hbox{ if and only if }\quad S(X)\leq S(Y).\eeq 
Equivalently,
\beq X\prec\prec  Y\hbox{ implies } S(X)<S(Y) \hbox{ and }X\sima Y \hbox{ implies } S(X)=S(Y).\label{1.9}\eeq
Moreover, $S$ is additive and extensive, i.e., 
\beq S(X,Y)=S(X)+S(Y)\quad\hbox{and} \quad S(\lambda X)=\lambda S(X).\eeq
The function $S$ is {unique} up to an affine change of scale, i.e., up to a replacement  $S(X)\to a_\Gamma S(X)+B(\Gamma)$ with constants $a_\Gamma$ and $B(\Gamma)$.
\end{itemize}
\end{thm}

\begin{proof}

\textit{Uniqueness: } 
Pick two {reference points} {$X_0\prec\prec X_1$} in {$\Gamma$} and
let {$X$} be an arbitrary state with {$X_0 \prec X\prec  X_1$}. 
For any entropy function {$S$} we have {$S(X_0)<S(X_1)$} and {$S(X_0)\leq S(X)\leq S(X_1)$} so there is a unique number {$\lambda$} between 0 and 1 such that
\beq S(X)=(1-\lambda)S(X_{0})+\lambda S(X_{1}). \eeq
By the required properties of entropy this is  {equivalent} to
\beq X\sima ((1-\lambda) X_{0},\lambda 
X_{1}).\eeq
Any other entropy function {$S'$}  also leads to (1.12) with {$\lambda$} replaced by some  {$\lambda'$}, but from the assumptions A1-A6 and {$X_0\prec\prec X_1$} a straightforward computation shows that 
(1.12) can hold for \textit{at most one} {$\lambda$}, i.e., {$\lambda=\lambda'$}. 

\textit{Existence: }From assumptions A1-A6 and $\mathrm{CH}$ one concludes that for {$X_0\prec X\prec X_1$} the following two numbers are {equal}:\begin{align}\lambda_{X,-}&:=\sup\{\lambda\,:\, ((1-\lambda)X_0,\lambda X_1)\prec X\}\\
\lambda_{X,+}&:=\inf\ \{\lambda\,:\, X\prec ((1-\lambda)X_0,\lambda X_1)\}.
\end{align}
Moreover, there is a unique {$\lambda_X$} at which that the sup and the inf are attained, and  \beq X\sima ((1-\lambda_X) X_{0},\lambda_X
X_{1}).\eeq 
The arguments leading to these conclusions are  detailed in Lemmas 2.2 and 2.3 in \cite{LY99}.

With the choice 
\beq S(X_{0})=0\quad \hbox{and}\quad S(X_{1})=1\eeq for some reference 
points $X_{0}\prec\prec X_{1}$, we have an  {\textbf explicit formula for the entropy} $S(X)=\lambda_X$:
\beq \boxed{S(X)=\sup\{\lambda\,:\, ((1-\lambda)X_0,\lambda X_1)\prec X\}}\label{1.17}\eeq
or equivalently
\beq\boxed{S(X)=\inf\ \{\lambda\,:\, X\prec ((1-\lambda)X_0,\lambda X_1)\}}\label{1.18}\eeq

{that uses  only the relation} {$\prec$}. \medskip

Any other choice of {$S(X_{0})$} and {$S(X_{1})$} leads to an affine transformation of the values of {$S$}, i.e, a replacement $S(X)\to a_\Gamma S(X)+B(\Gamma)$ with constants $a_\Gamma$ and $B(\Gamma)$.
\end{proof}

\rem The numbers $\lambda$  in Eqs. (1.17) and (1.18)  are  not restricted to $[0,1]$ if we use the convention that $(X,-Y)\prec Z$ means the same as $X\prec (Y,Z)$. In fact, if $X\prec\prec X_0$, or $X_1\prec\prec X$ then $\lambda$ is $<0$ or $>1$ respectively. 

\rem[] Figure \ref{fig1}  illustrates schematically Eqs. (1.17)  and (1.18). The entropy $S(X)$ can by Eq. (1.17)  intuitively be regarded as the maximum amount of substance in the \lq\lq high entropy\rq\rq\ state $X_1$ that can be changed adiabatically to the state $X$ with the aid of a complementary amount of substance in the \lq\lq low entropy\rq\rq\  state $X_0$. Equivalently, by Eq. (1.18)  it is the minimum amount of substance the state $X_1$ that arises when a complementary amount of $X$ is changed adiabatically to the state $X_0$.

\begin{figure}
\centering
\begin{subfigure}{.5\textwidth}
 \hskip-7.5cm
  \includegraphics[width=14cm]{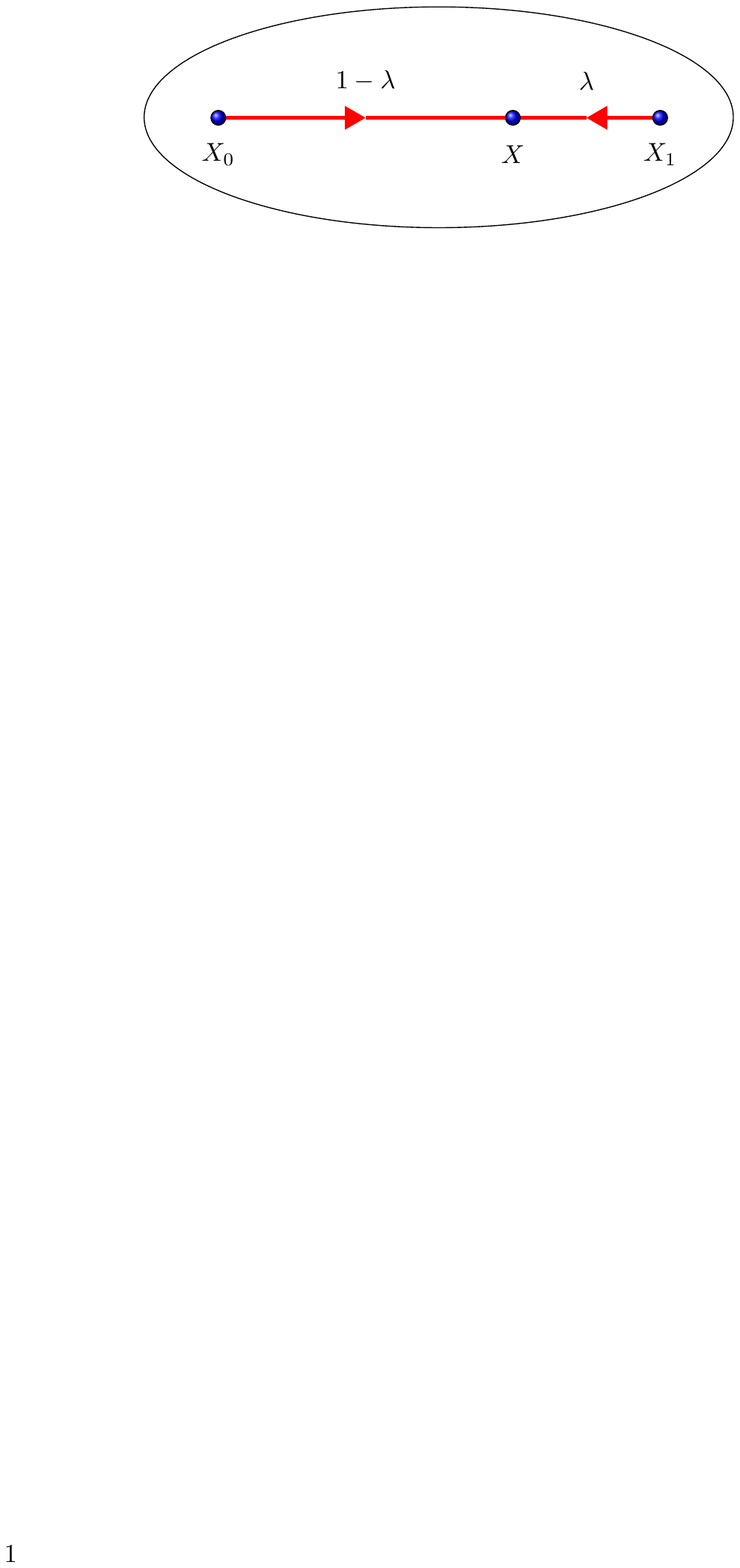}
\end{subfigure}%
\begin{subfigure}{.5\textwidth}
\hskip-8.5cm
  \centering
  \includegraphics[width=14cm]{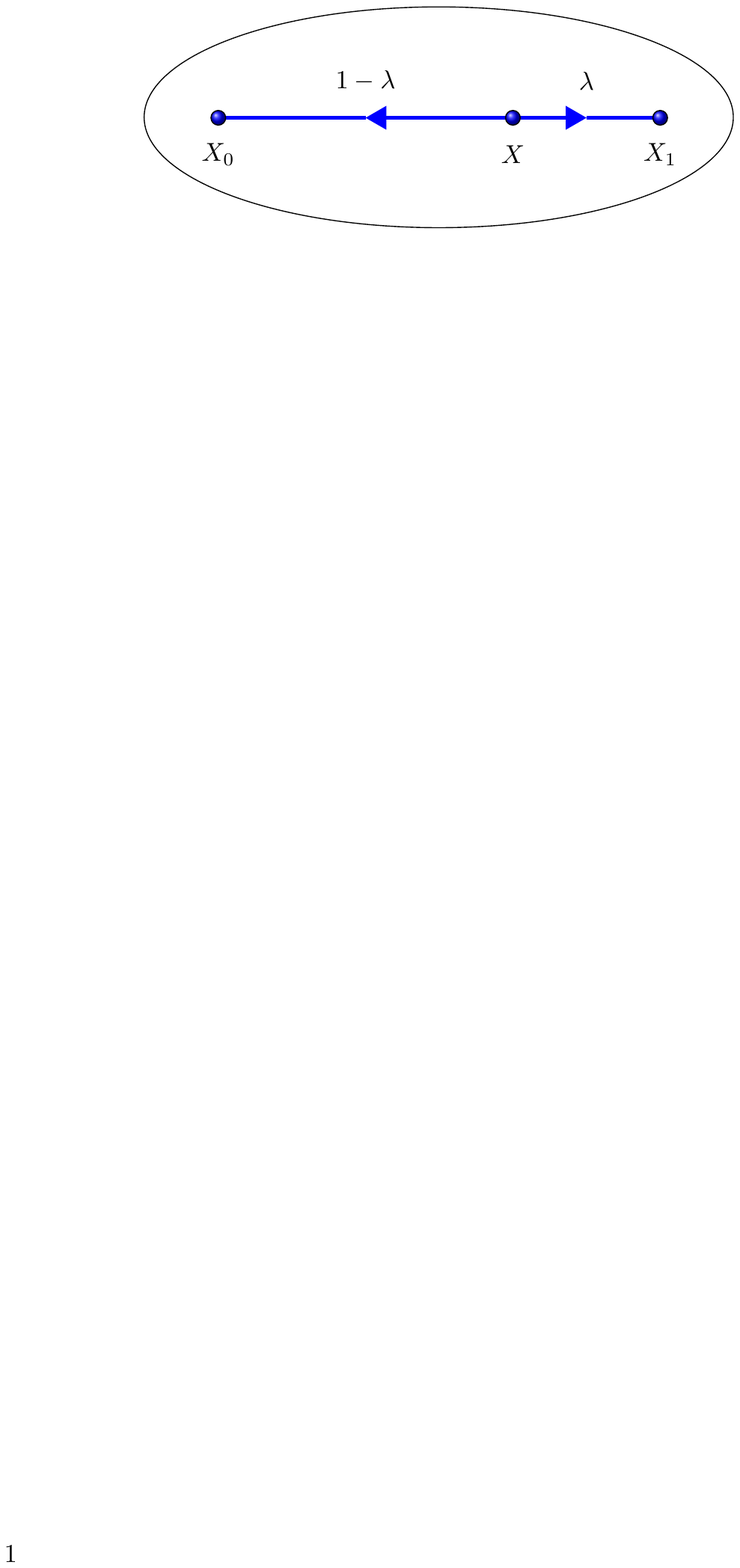}
\end{subfigure}\vskip-13cm
\caption{\small Definition of entropy, cf. Eqs. \eqref{1.17} and \eqref{1.18}. The left figure illustrates the processes employed for the definition by Eq. \eqref{1.17},  the right figure the analogous processes for \eqref{1.18}.}
\label{fig1}
\end{figure}

\rem
Form the proof it is evident that comparability of all states in 
$(1-\lambda)\Gamma\times \lambda \Gamma$ is essential,  not only of those in $\Gamma$.  In particular (by A4) all states in $\Gamma\times\Gamma$ must be comparable.\medskip

The following simple example shows that this is a nontrivial requirement: $\Gamma$ consists of the positive real numbers $\mathbb R^+$ and the relation $\prec$ on $\Gamma$ is the usual order on $\mathbb R^+$. If the order on $\Gamma\times\Gamma$ is defined by $(x,y)\prec (x',y')$ if and only if $x\leq x'$ and $y\leq y'$, then it is clear that e.g. $(1, 2)$ and $(2,1)$ are not comparable.\medskip

One  important point about the general definition of entropy remains to be discussed. By Theorem 1 the entropy for a system $\Gamma$ is only unique up to constants $a_\Gamma$ and $B(\Gamma)$ which are arbitrary as long as one considers only state changes within scaled copies of $\Gamma$. In order to fix the entropy on products of different systems maintaining additivity and extensivity, it is necessary to choose the free constants in a consistent way. Theorem 2.5 in \cite{LY99} states that this can always be achieved, provided CH has been proved to holds also for products  of different systems. The idea for the proof is to pick one  system $\Gamma_0$ as a standard reference and consider products of other systems with $\Gamma_0$. The precise statement is a follows:

\begin{thm}[\textbf{Consistent entropy scales} ]Assume that CH holds for all compound systems. For each system $\Gamma$ (which can be a compound system)) let $S_\Gamma$ be some definite entropy function for $\Gamma$ in the sense of Theorem 1. Then there are constants $a_\Gamma$ and $B(\Gamma)$ such the the function $S$ defined for all state of all systems by 
\beq S(X)=a_\Gamma S_\Gamma(X)+B(\Gamma)\label{1.19}\eeq
for $X\in\Gamma$ is additive and extensive and characterizes the relation $\prec$ on all compound systems in the sense of \eqref{1.9}.
\end{thm}

It should be emphasized  the knowledge that an \textit{additive} entropy $S$ characterizing the relation on the product of two state spaces, $\Gamma_1\times \Gamma_2$, conveys much information: In order to deduce that 
$(X,Y)\prec(X',Y')$ it is only necessary to know that the sum $S(X)+S(Y)$ is no less than the sum $S(X')+S(Y')$, and this can hold even if one of the entropies, $S(X)$ or $S(Y)$, decreases, provided the decrease is compensated by an increase of the other entropy.



\section{Simple systems}
 Simple systems are the building blocks of thermodynamics. A simple system contains a fixed amount of matter of fixed chemical composition and its states are parametrized by \textit{one} energy coordinate $U$  and one or more work coordinates, denoted $V=(V_1,\dots, V_n)$ because the volume is a typical one.  For a  fluid or a gas in a container the volume is usually the only work coordinate.  Simple system can, however,  be more complex. Several containers of a gas connected by copper threads, so that the containers can freely exchange energy, form together a simple system with the volumina $V_i$ of the  individual containers as the work coordinates and  the sum of the energies as the energy coordinate. For a solid, components of the strain tensors may serve as work coordinates and for a magnet, the magnetization takes this role.
 
 The state space of a simple system can in a natural way be regarded as a subset of $\mathbb R^{n+1}$ with $n\geq 1$ the number of work coordinates. We now put forward an axiom about convex combinations of states:
 
 {\textbf {A7}.} \textit{Convex combinations}. The state space $\Gamma$ of a simple system is an open convex subset of $\mathbb R^{n+1}$, and the formation of a convex combination of states is an adiabatic operation. More precisely, if $X, Y\in\Gamma$ and $t\in [0,1]$ then
 \beq (tX,(1-t)Y)\prec tX+(1-t)Y\eeq
Note that  the state on the left hand side is in  $t\Gamma\times (1-t)\Gamma$ while right hand side is a state  in $\Gamma$.

This axiom is well motivated for gases and liquids, where the convex combination can be operationally achieved by removing a separating wall between the states in $t\Gamma$ and $(1-t)\Gamma$ respectively and waiting for a new equilibrium state to establish itself. For solids the axiom may be less obvious, but it is experimentally testable since it is \textit{equivalent to concavity of the entropy} as a function of $U$ and the work coordinates. Further discussion can be found on p. 33 in \cite{LY99}. The axiom makes sense for any state space $\Gamma$ with a convex structure, hence  it can be  regarded as a general axiom like the previous axioms A1-A6. It implies in particular that the \textit{forward sector} of a state $X\in \Gamma$, defined as
\beq A_X:=\{Y\in \Gamma\, :\, X\prec Y\},\eeq
is a convex subset of $\Gamma$.\medskip

The following new axioms are specific for simple systems. 
\begin{itemize}
\item[\textbf {S1.}]  \textit{Existence of irreversible state changes}. For every $X\in \Gamma$ there is a $Y\in\Gamma$ such that $X\prec\prec Y$, i.e., $X\prec Y$ but $Y\not\prec X$.
\item[{\textbf S2.}]  \textit{Tangent planes.} For every $X$ the forward sector $A_X$ has a \textit{unique} tangent plane at $X$ and its normal is not orthogonal to the $U$-axis. The tangent plane is assumed to be a locally Lipschitz continuous function of $X$ in the sense explained below.
\item[{\textbf S3.}] \textit{Connectedness of the boundary}. The boundary $\partial A_X$ of the forward sector $A_X$ is a connected set.
\end{itemize}

\rem[] Axiom S1 is in our framework the basic assumption that makes contact to the traditional formulations of the second law of thermodynamics. As in the classical versions by Carnot, Clausius and 
Kelvin it claims the \textit{impossibility} of certain processes. Together with the convexity axiom A7 it is, in fact, \textit{equivalent} to  Carath\'{e}ordory's formulation of the second law \cite{Cara09}: In every neighbourhood of every state there are states that are states that are not adiabatically accessible from it. Axiom S1 is at first sight a much weaker requirement, because for every $X$ it makes only a claim about one $Y$ which might a priori be arbitrarily far from $X$. The convexity axiom A7, however,  together with the other axions A1-A6 implies in fact an equivalence of both statements, see Theorem 2.9 in \cite{LY99}.

\rem[] Carath\'{e}odory's principle implies in particular that $X$ is on the boundary $\partial A_X$ of the convex forward sector $A_X$ which, by the convexity axiom A7, lies on  one side of any supporting hyperplane at $X$. Axiom S2 requires that  this  hyperplane is \textit{unique}, i.e., $A_X$ has a tangent plane at $X$. (In other words, the boundary has no curbs.) Moreover, the normal of this tangent plane is by assumption not orthogonal to the $U$-axis so for $X=(U^0,V^0)$ the plane is given by an equation
\beq U-U^0+\sum P_i(X)(V_i-V^0_i)=0
\eeq
with locally Lipschitz continuous functions $P_i$ which are the generalized \textit{pressures} corresponding to the work coordinates $V_i$. Lipschitz continuity and the connectedness axiom S3 lead to the conclusion that  the coupled differential equations
\beq \partial U/\partial V_j(V)=-P_j(U(V), V)\quad\hbox{for $j=1,\dots,n$}\label{1.23}\eeq
have a unique solution with $U(V^0)=U^0$. This solution describes the adiabat $\partial A_X$ as a submanifold in $\Gamma$.

\rem[] In conventional thermodynamic notion Eq. \eqref{1.23} can be written
\beq dU+P\cdot dV=0.\label{1.24}\eeq
The differential form on the left hand side is in textbooks often called "infinitesimal heat" and  denoted by $\delta Q$ but we shall not use this terminology.  The fact that Eq. \eqref{1.23} defines an $n$ dimensional submanifold, parametrized as $V\to U(V)$, through every $X=(U^0,V^0)$, is a consequence of Axiom S2. In mathematical terminology it means that that the differential form $\delta Q$ has an integrating factor.\medskip

\begin{figure}[ht]
\centering
\includegraphics[width=5.5cm]{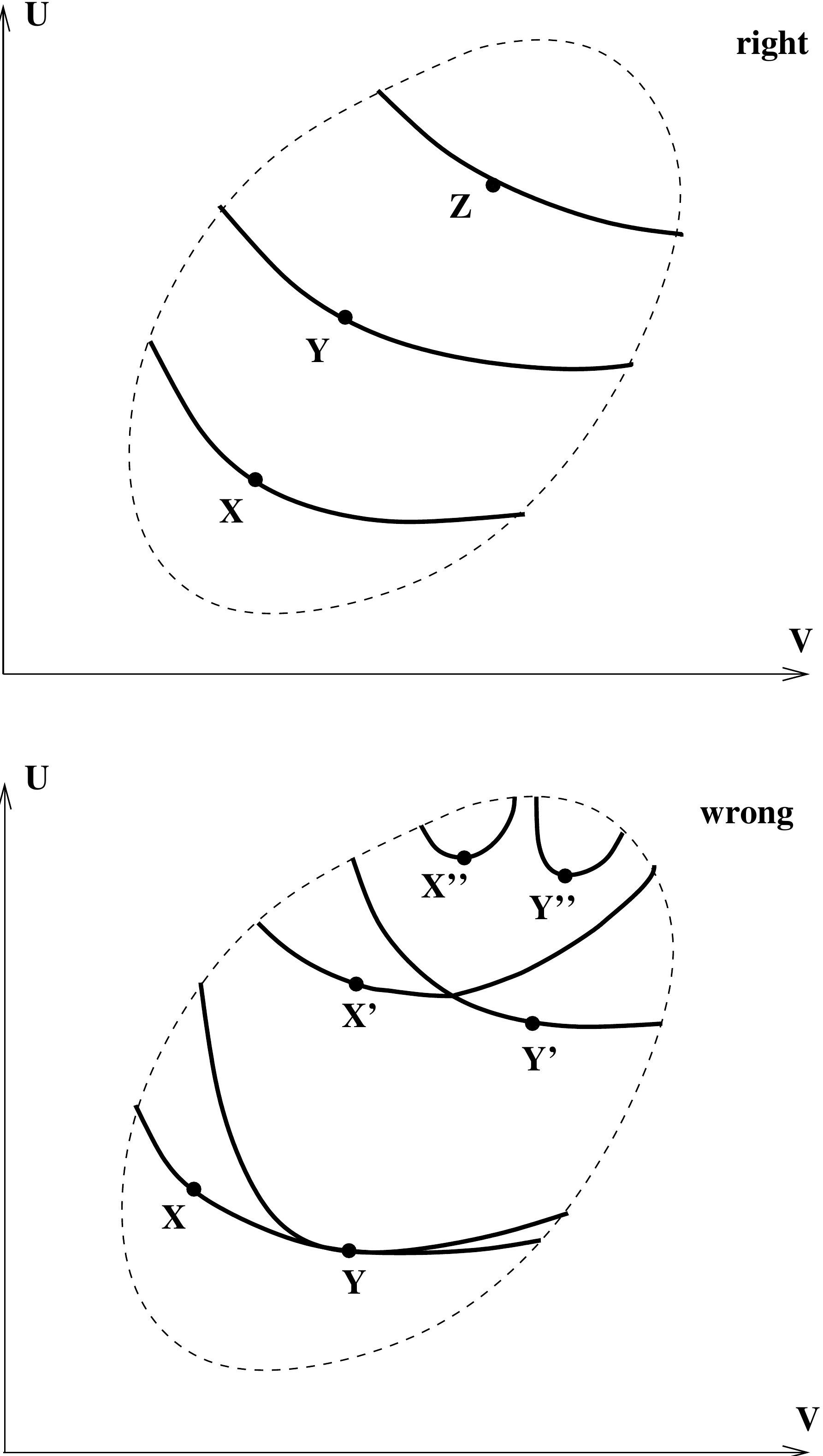}
\caption{The upper figure illustrates that forward sectors in the state space of a simple system (enclosed within the dashed curve) are nested.  The adiabat 
$\partial A_X$ is the hypersurface containing $X$, and the forward sector $A_X$ is the region above it. Likewise for the states $Y$ and $Z$. The lower figures shows what in principle could go wrong, but doesn't because of our axioms, in particular S2.\label{fig2}}
\end{figure}

The axioms A1-A7 and S1-S3 stated so far imply comparability for all pairs of state of a simple system. More precisley, the following theorem hods: 
\begin{thm}[\textbf {Comparability within a simple system}] If $X$ and $Y$ are states of the same simple system, then either $X\prec Y$ or $Y\prec X$ (or both). Moreover, $X\sima Y$ if and only if $Y\in \partial A_X$.

\end{thm}

In geometrical terms the theorem says ist that the forward sectors of a simple systems are nested, see fig. \ref{fig2}. 
The proof of this theorem is  given in  Section 3 in \cite{LY99}.  It requires more mathematical  effort than the proof of the Theorems 1 and 2 in the last Section, where only the general axioms A1-A6 (plus CH) were used.

\section{Thermal contact and absolute temperature}

While Theorem 3 establishes  comparability of states for a simple system this is not yet sufficient for the definition of an additive and extensive entropy by Theorem 1.  What is needed is that comparability holds also within scaled products of such systems. Here the the concept of \textit{thermal contact } and \textit{thermal equilibrium} enters our analysis. The basic idea is that by joining two systems with a "copper thread" so that they can freely exchange energy at fixed work coordinates we can form a new system system to which the analysis of the previous section applies. This idea is formalized in new axioms.

\begin{itemize}
\item[\textbf {T1.}]  \textit{Forming a thermal join}. For any two simple systems with state spaces $\Gamma_1$ and $\Gamma_2$ there is another simple system, called the \textit{thermal join} of the two spaces with state space
\beq \Delta_{12}=\{(U,V_1, V_2)\ :\  U=U_1+U_2 \hbox{ with } (U_1,V_1)\in \Gamma_1, (U_2,V_2)\in \Gamma_2\}.\eeq
Moreover, the formation of the adiabatic join is an adiabatic operation:
\beq \Gamma_1\times \Gamma_2\ni ((U_1,V_1), (U_2,V_2))\prec (U_1+U_2, V_1,V_2)\in \Delta_{12}\label{1.26}\eeq
 \item[\textbf {T2.}]  \textit{Splitting  a thermal join}. For any $(U,V_1, V_2)\in \Delta_{12}$ there is at least one pair of states $(U_1,V_1)\in \Gamma_1, (U_2,V_2)\in \Gamma_2$ with $U_1+U_2=U$ and such that
\beq (U,V_1, V_2)\sima ((U_1,V_1), (U_2,V_2))\label{17}\eeq

\textbf{Definition.} If \eqref{17} holds we say that the states $X$ and $Y$ on the right hand side are in \textit{thermal equilibrium} and write \beq X\simt Y.\eeq

\item[\textbf {T3.}] \textit{Zeroth law of thermodynamics.} If $X,Y,Z$ are states of three, in general different, simple systems then $X\simt Y$ and $Y\simt Z$ implies $X\simt Z$.
\end{itemize}
The zeroth law says that states in thermal equilibrium fall into equivalence classes which are conventionally labelled by some empirical temperature scale. We shall not use that concept at this point because we are soon to define the absolute temperature scale below,

These axioms are all all essential for our proof of the entropy principle for products of simple systems. They are still not enough, however. A further axiom, called \textit{transversality} requires that \textit{isotherms}, i.e., equivalence classes w.r.t. the relation $\simt$ must contain states on both sides of the adiabats $\partial A_X$ which are equivalence classes w.r.t. the relation $\sima$.

\begin{itemize}
\item[\textbf {T4.}] \textit{Transversality.} If $\Gamma$ is the state space of a simple system and $X\in\Gamma$, then there exist states $X_0, X_1\in\Gamma$ such that $X_0\simt X_1$ and $X_0\prec\prec X\prec\prec X_1$.
\end{itemize}

The significance of this axiom can be seen by considering a simple system $\Gamma$ where the pressure is identically zero  and the relation $\prec$ therefore independent of the work coordinates: $(U,V)\prec (U',V')$ if and only if $U\leq U'$. The thermal join $\Delta$ of $\Gamma\times \Gamma$ is again a system of the same type and $(U_1,V_1)\simt (U_2, V_2)$ means that $U_1=U_2=\hbox{$\frac 12$} (U_1+U_2)$. The isotherms coincide therefore with the adiabats. The comparison hypothesis for $\Gamma\times\Gamma$ is violated as can be seen from Fig. \ref{fig3}.\footnote{For a further discussion of this point and the possibility to rescue CH by coupling a system violating T4 to a "normal" system see Section 4.3 in \cite{LY99}.}
\begin{figure}[ht]
\centering
\includegraphics[width=6cm]{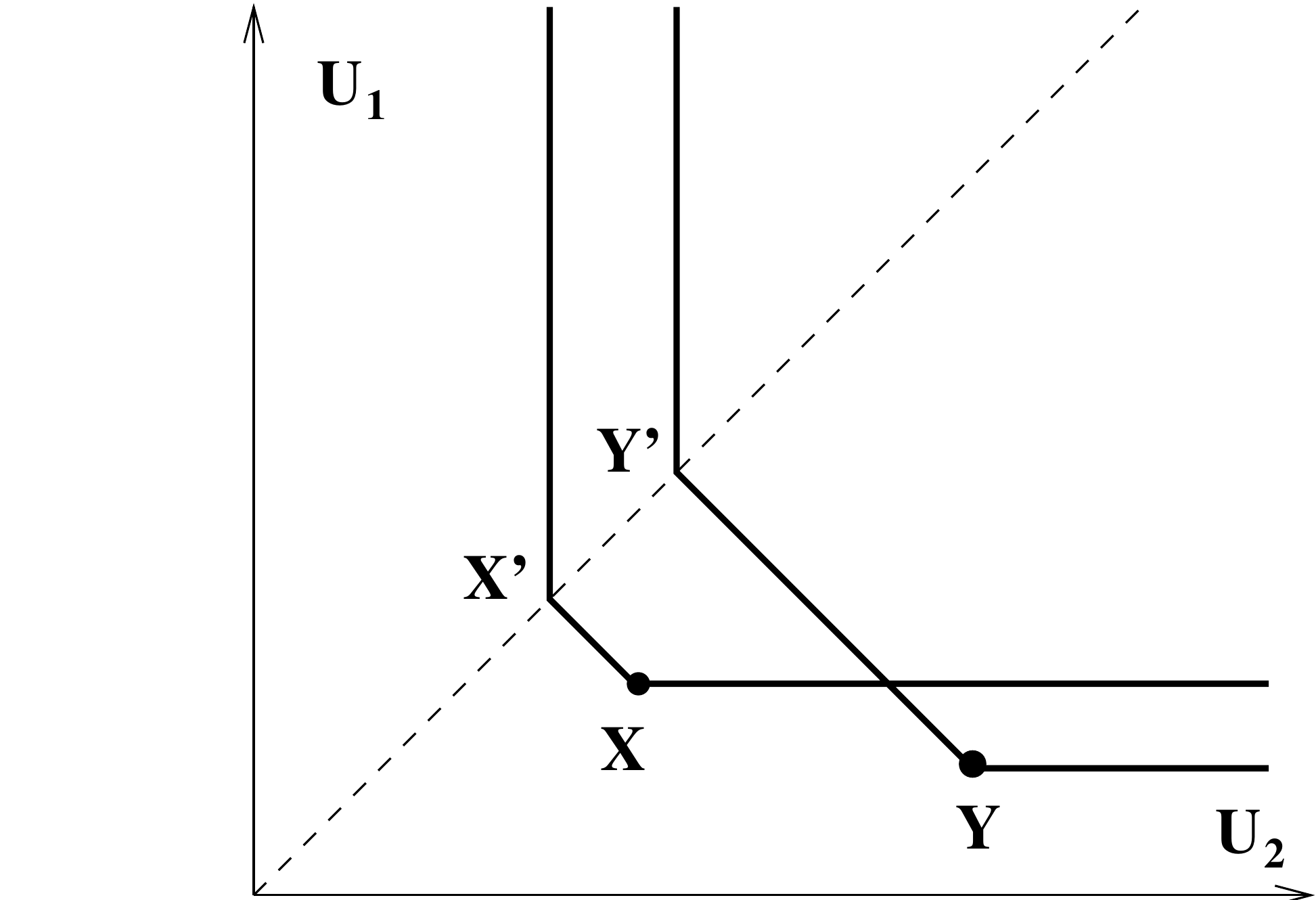}
\caption{The figure illustrates what can happen if the transversality axiom A4 does not hold. If the pressure is independent of $V$ isotherms and adiabats for a simple system $\Gamma$ coincide and the relation $\prec$ depends only on the energy. The figure shows  the state space of $\Gamma\times\Gamma$, ignoring the volumina, and two states, $X$ and $Y$, in $\Gamma\times\Gamma$. The thermal join $\Delta$ coreponds to the dashed line on the diagonal, $X'$ and $Y'$ are the projections the two states onto $\Delta$ as in \eqref{1.26}. It is evident from the figure that that $X$ and $Y$ are not comparable in $\Gamma\times\Gamma$.} {\label{fig3}}. 
\end{figure}

A last thermal axiom is technical and can possibly be eliminated. Intuitively it can be thought of saying that all systems have the same range of temperatures. (A rigorous definition of "temperature" is given in Eq. \eqref{1.30} below.)

\begin{itemize}
\item[\textbf {T5.}] \textit{Universal temperature range}. Let $\Gamma_1$ and $\Gamma_2$ be state spaces of simple systems.  For every $X\in \Gamma_1$ there is a $Y\in \Gamma_2$ such that $X\simt Y$. Moreover, such a $Y$ exists with any prescribed value of the work coordinates for $\Gamma_2$. \end{itemize}

The thermal axioms axioms in conjunction with the previous axioms A1-A7 and S1-S3 are sufficient to establish the comparison hypothesis for all scaled products of simple systems and hence an additive and extensive entropy by the procedure of Section 1. The basic idea is to use that the thermal join $\Delta_{12}$ of two simple systems is a simple system for which comparability of all states holds by Theorem 3. The implementation of this idea is not entirely simple and it uses all the previous axioms, in particular the zeroth law, T3, and the transversality axiom T4. Also concavity of entropy as a function of $U$ which follows from A7 is important. 

The full proof of the CH is presented in Section 4.2 pp 59-64 in \cite{LY99} 
(see also pp. 109--113 in \cite{LY01}).
 It is done in two steps. First, one considers multiple scaled products of a single simple system $\Gamma$. Here a key point is that if $X, X_0, X_1\in \Gamma$ are as in Axiom T4, then the states $((1-\lambda) X_0, \lambda X_1)$ and $((1-\lambda)X,\lambda X)$ are adiabatically equivalent to a state of the same simple system and hence comparable. In a second step one considers products of different simple systems. This case is more complicated and the zeroth law, which was not needed in the first step, is used here. A crucial lemma is the following:
\begin{lem}[\textbf{Existence of calibrators}] For any pair of state spaces $\Gamma_1,\Gamma_2$ of simple systems there exist states $X_0,X_1\in \Gamma_1$ and $Y_0,Y_1\in \Gamma_2$ such that $X_0\prec\prec X_1$, $Y_0\prec\prec Y_1$ and $(X_0,Y_1)\sima (X_1,Y_0)$.
\end{lem} 
Note that this lemma is trivial in the case that $\Gamma_1=\Gamma_2=\Gamma$ because  one can then  simply take $Y_0=X_0$ and $Y_1=X_1$. For different systems the statement is not obvious and a proof is required. It is given on p. 63 in \cite{LY99}. The reason for the name of the lemma is that if $\Gamma_1\neq \Gamma_2$ it allows a calibration of the so far undetermined multiplicative constants for the entropies $S_1$ and $S_2$ of these systems to ensure additivity.
This is done by requiring  
\beq S_1(X_0)+S_2(Y_1)=S_1(X_1)+S_2(Y_0).\eeq
\medskip
Using the lemma one can conclude that the prerequisites for Theorem 2 are fulfilled and we obtain

\begin{thm}[\textbf{Entropy principle in products of simple systems}]
The comparison hypothesis is valid in arbitrary scaled products of 
simple systems. Hence the relation of adiabatic accessibility   in such state spaces 
is characterized (in the sense of  Eq. (1.9)) by an  addi\-tive and extensive entropy, $S$, which is  unique up to an overall multiplicative constant and one 
additive constant for each  simple system under consideration.
\end{thm}

The uniqueness is most important.  It means that the entropy defined by the abstract formulas of Section 1 can be determined in the same way as  in standard thermodynamics by integration involving measurable quantities like heat capacities, compressibilities etc. 
But first we must define \textit{temperature}! In our approach it comes at the end of the analysis as a corollary to entropy and not the beginning as in traditional treatments. Using strict concavity of entropy, implied by A7  and S1, the regularity assumption S2 for the pressure, and the transversality axiom T4 the following theorem is proved in Section 5 in \cite{LY99}:

\begin{thm}[\textbf{Entropy defines temperature}] The entropy $S$ is a concave and continuously differentiable function on the state space of a simple system and it is nowhere locally constant. If the function $T$ is defined by
\beq\frac 1T:=\left(\frac {\partial S} {\partial U}\right)_{\mathrm V}\label{1.30}\eeq
then $T>0$ and $T$ characterizes the relation \hbox{\rm $\simt $}  in the sense that \hbox{\rm $X\simt Y$} if and only if $T(X)=T(Y)$. Moreover, if two systems are brought into thermal contact the energy flows from the the system with the higher $T$ to the one with the lower $T$.
\end{thm}

\rem[\textbf{Sign of the temperature}] In our framework temperature has one sign which is positive by our convention that the forward sectors $A_X$ point in the direction of increasing $U$. Our axioms, in particular S1 together with A7, do not allow to consider simultaneously systems with different signs of temperature. See pp. 47-48 in \cite{LY99}. For other viewpoints see \cite{Lavis} and references cited therein.

\rem[\textbf{Practical determination of absolute temperature and  entropy}]  By \eqref{1.30} we have
\beq dS=\frac 1T dU+\frac  P T d V.\label{1.31}\eeq
The fact that the right hand side of \eqref{1.31} is a total differential, so that the derivative of $1/T$ with respect to $V$ is the same as the derivative of $P/T$ with respect to $U$,  has many important consequences,  as discussed in standard thermodynamic texts.  An example is a formula due to Max Planck \cite{P} pp.\ 134-135, which relates the absolute temperature $T$ to any arbitrary  \textit{empirical temperature scale} $\Theta$, i.e, any continuous function $\Theta$ with $\Theta(X)=\Theta (Y)$ if and only if $X\simt Y$:
  \begin{equation}\label{5}
   T(\Theta)=T_0 \exp
   \left(\,\,
   \int_{\Theta_0}^\Theta{\frac{\left(\frac{\partial P}{\partial \Theta'}\right)_{\!V}}{P+\left(\frac{\partial U}{\partial V}\right)_{\!\Theta'}}\;d\Theta'}
  \right).
  \end{equation}
  This is proved by solving the differential equation resulting from the integrability condition for the differential form $dS$, using that constant $T$ means the same as constant $\Theta$.
  It is remarkable that the integral on the right-hand side depends only on the temperature although the terms in the integrand depend in general also on $U$ and $V$.
  
Absolute temperature can thus be determined from directly measurable quantities. The same holds for entropy:
\beq S(X)=S(X_0)+\int_{X_0\to X} \left(\frac 1T dU+\frac  P T d V\right)\eeq
where the integral can be taken over any convenient path in state space leading from $X_0$ to $X$ because  $dS$ is a total differential. 

As a final remark on temperature we note that isotherms, i.e. the sets where temperature is constant, can be more complex geometrically than the adiabats. While the latter are smooth boundaries of convex sets the isotherms have no such property in general. They need not even have co-dimension one, cf. Fig. \ref{fig4}   which shows schematically isotherms in the $(U,V)$ plane around the triple point of a  substance with three phases.

\begin{figure}[ht]
\centering
\includegraphics[width=6cm]{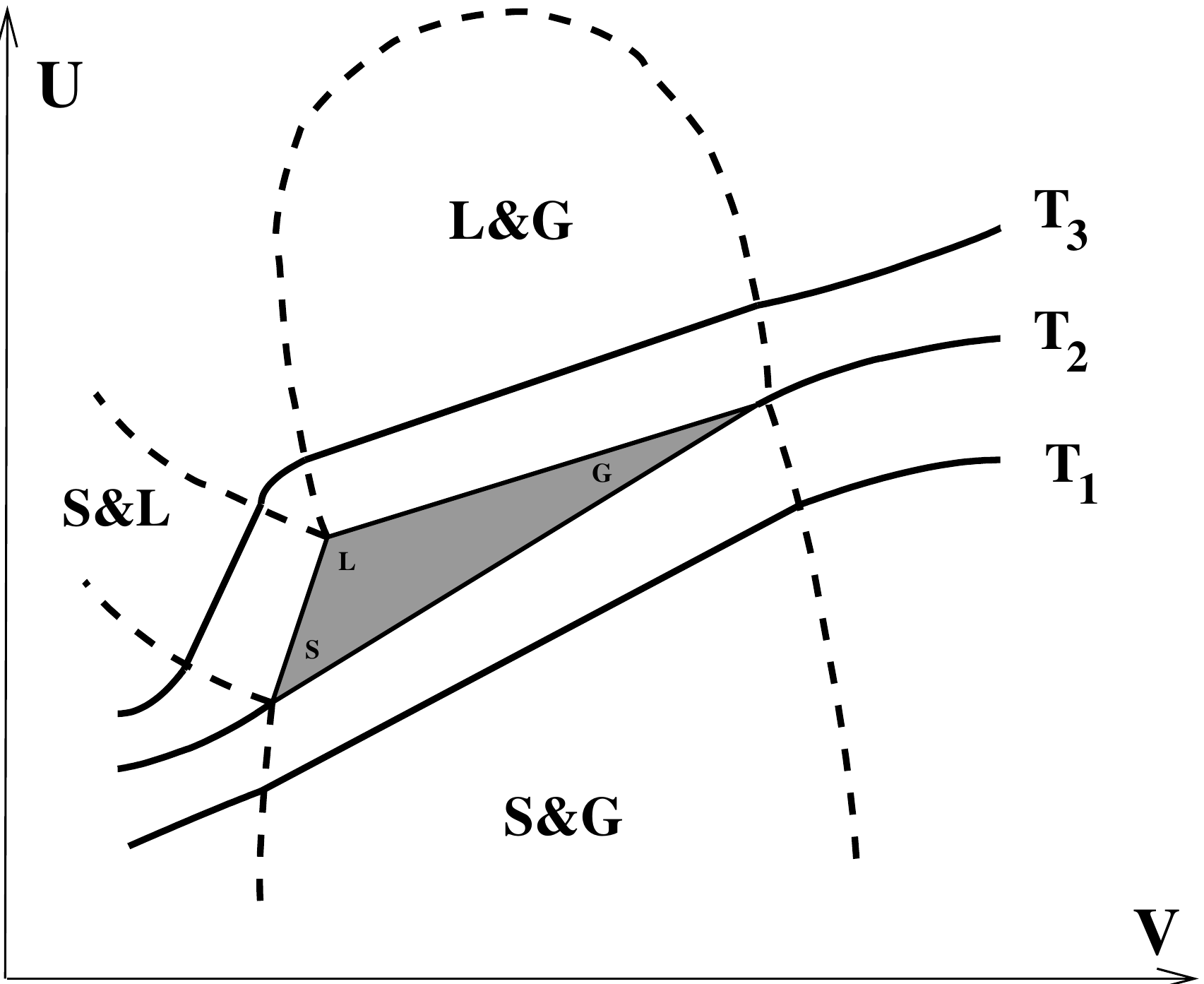}
\caption{The figure (not to scale) shows isotherms in the $(U,V)$ plane near the triple point of a system with three phases (solid S, liquid L, gas G). In the triple point region the temperature is constant which shows that isotherms need not have codimension 1.}{\label{fig4}}
\end{figure}

\section{Mixing and chemical reactions} 

Theorem 4 is the culmination of our analysis of the entropy principle under the assumptions that systems have fixed chemical compositions so that mixing and chemical reactions are excluded. For a wide range of applications of thermodynamics this is fully satisfactory. Nevertheless, there are important situations, in particular in chemistry, where mixing of different substances or chemical reactions can be regarded as adiabatic state changes in the sense of the definition in the Introduction. Here the relation $\prec$ does not conserve individual systems and this possibility is not covered by our theory as described so far.

Mathematically  one can formulate the problem as the question of determining the additive constants $B(\Gamma)$ in Theorem 2, Eq. \eqref{1.19},  in a way that is consistent with additivity of entropy and the requirement that $X\prec Y$ implies $S(X)\leq S(Y)$ even if $X$ and $Y$ belong to different state spaces. If one had at one's disposal \textit{semipermeable membranes} and \textit{van t'Hofft boxes}, which allow adiabatically reversible mixing processes and chemical reactions, this would not be difficult and is discussed in standard textbooks, e.g. in \cite{Pau00}, Ch. 3, or \cite{F}, Ch. 6. As noted in \cite{F}, p. 101, however, \lq\lq in reality no ideal semipermeable membranes exist\rq\rq. In  \cite{LY99} we proposed another way without assuming the existence of such idealized objects. The outcome of the analysis in Sec. 6 in \cite{LY99} is the following theorem. It requires one final axiom, denoted \textbf M,  which is discussed further below.

\begin{thm}[\textbf{Universal entropy}]  The additive entropy constants of all systems can be calibrated in such a way that the entropy is additive and extensive and $X\prec Y$ implies $S(X)\leq S(Y)$ even if $X$ and $Y$ do not belong to the same state space.
\end{thm}

A basic input for the proof of this theorem the fact that as a consequence of Theorem 4 the multiplicative constants $a_\Gamma$ can be determined in such a way that for any compound space $\Gamma_1\times\Gamma_2\times \Gamma_3,\dots$ and $X_i, Y_i\in \Gamma_i$ the relation 
$(X_1,X_2,\dots)\prec (Y_1,Y_2,\dots)$ holds if and only if $\sum_i S(X_i)\leq \sum_iS(Y_i)$. The additive  constants in \eqref{1.19} are unimportant here because their sum on both sides is the same. The important point is that this hods even if the process taking $(X_1,X_2,\dots)$ to $(Y_1,Y_2,\dots)$ takes in intermediate steps one system to another, provided the total compound systems is the same a the beginning and the end. This information can be turned into an inequality for the maximal \lq\lq mismatch\rq\rq\ between the entropy constants for different spaces $\Gamma$ and $\Gamma'$, expressed through a function $F(\Gamma, \Gamma)$ with values in $\mathbb R\cup\{-\infty\}\cup\{+\infty\}$ which satisfies for $X\in\Gamma$, $Y\in\Gamma'$ the condition
\beq  X\prec Y\hbox{ if and only if } S_\Gamma(X)+F(\Gamma, \Gamma')\leq S_{\Gamma'}(Y).\label{1.34}\eeq
The precise definition of $F(\Gamma,\Gamma')$ and the proof of Eq. \eqref{1.34} is given on p. 82 in \cite{LY99}, see also p. 12--121 in \cite{LY01}. The definition of $F$ involves considering sums of entropy differences for chains of state spaces which lead from $\Gamma\times\Gamma_0$ to $\Gamma'\times \Gamma_0$ where $\Gamma_0$ is an auxiliary state space that plays the role of a \lq catalyst\rq.  Given Eq.\ \eqref{1.34} the proof of Theorem 6 amounts to determining the entropy constants $B(\Gamma)$ in such a way that
\beq -F(\Gamma',\Gamma)\leq B(\Gamma)-B(\Gamma')\leq F(\Gamma,\Gamma')\eeq and this is done by appealing to certain subadditivity properties of the function $F(\Gamma,\Gamma')$ and using the Hahn-Banach Theorem. In order to obtain finite constants $B(\Gamma)$, however, one final axioms is needed to exclude that $F(\Gamma,\Gamma')=-\infty$. We say that a state space $\Gamma$ is \textit{connected to another state space} $\Gamma'$ if $F(\Gamma,\Gamma')<\infty$.  The meaning ist that there is a chain of states in intermediate state spaces, beginning in $\Gamma\times \Gamma_0$ and ending in $\Gamma'\times\Gamma_0$.
such that the sum of the entropy jumps between the steps is finite. 
\begin{itemize}
\item[\textbf {M.}] \textit{Absence of sinks}. If $\Gamma$ is connected to $\Gamma'$ then $\Gamma'$ is connected to $\Gamma$.
\end{itemize}
This axiom excludes in particular the logical possibility that a substance can be synthesized from the chemical elements but it is impossible to recover the elements back.

Theorem 6 is not quite the optimal result that one could prove if semipermeable membranes were at our disposal. Then we could claim the converse statement, namely that  $S(X)\leq S(Y)$ implies $X\prec Y$ if both states contain the same amount of each of the 92 chemical elements. The entropy constants would indeed be fixed for all systems as soon as they have been chosen (arbitrarily) for each element. Without invoking semipermeable membranes one can, however,  in principle test experimentally, whether such a choice is sufficient to remove all possible nonuniquneess. Indeed, nonuniqueness would manifest itself in a genuine gap
\beq -F(\Gamma',\Gamma)<F(\Gamma,\Gamma')\eeq
for some systems $\Gamma$, $\Gamma'$  although both sides are finite. Such a situation is not, as far as we know, realized in nature.

\section{Conclusions} In the preceding sections I have sumarized the analysis of Elliott and myself of the entropy principle for thermodynamic equilibrium states  that was carried out in \cite{LY99}, based on the 16 axioms A1-A7, S1-S3, T1-T5 and M. That analysis has aged remarkably well over the more than 20 years since it was first presented. In 2013 and 2014 we extended it in two ways. In \cite{LY13} we considered the question what can be said about  nonequilibrium states with our methods, and  in \cite{LY14} we included systems for which our assumptions about scaling do not hold. In both cases the analysis  was based on some relaxations of the general axioms A1-A6 while the mathematically more sophisticated analysis using the rest of the axioms,  does not have a counterpart in these papers. The papers bring into focus the pivotal role of the comparison hypothesis for the results in \cite{LY99}, because that hypothesis can in general not been expected to hold in nonequilibrium situations.  Also for in the general setting of \cite{LY14} it must be postulated separately and we did not derive it from other axioms as in \cite{LY99}. I shall not discuss further details here but only remark that in the general setting of \cite{LY13} or \cite{LY14} the entropy need not have the uniqueness and additivity  properties as equilibrium entropy has, but one must in general expect a whole family of entropies, lying between two extremes denoted in the papers by by $S_-$ and $S_+$. Note in this context that the two formulas \eqref{1.17} and \eqref{1.18} give the same result only because comparability was assumed.

Finally, I want to stress that our road to a mathematically rigorous foundation for  thermodynamics, although in our opinion a rather direct one, is  not the only route. As mentioned in the Introduction, Section 1.2 in \cite{LY99} contains an overview of many alternative approaches that were on the market at the date of the publication of the paper. More recent examples are  \cite{Abou} and \cite{Kam}, and in the end it is a matter of anyone's taste which she or he prefers.

\section*{Acknowledgements} In the acknowledgements of \cite{LY99} we mentioned many people who were helpful to us when we wrote that paper and I am happy to repeat their names here (in alphabetical order): Fredrick Almgren, Thor Bak, Bernhard Baumgartner, Perluigi Contucci, Robin Giles, Roy Jackson, Martin Kruscal, Jan Philip Solovej and  John C. Wheeler. Last, but not least, I express my deep gratitude to Elliott for  his friendship and wonderful collaborations on many subjects over more than three decades. My contribution \cite{ Y} to the special volume of Journal of Statistical Physics on the occasion of Elliott's seventieth birthday contains some recollections from the time when we worked on the paper \cite{LY99}.

\begin{figure}
\centering
\begin{subfigure}{.5\textwidth}
 \hskip-7.5cm
  \includegraphics[width=14cm]{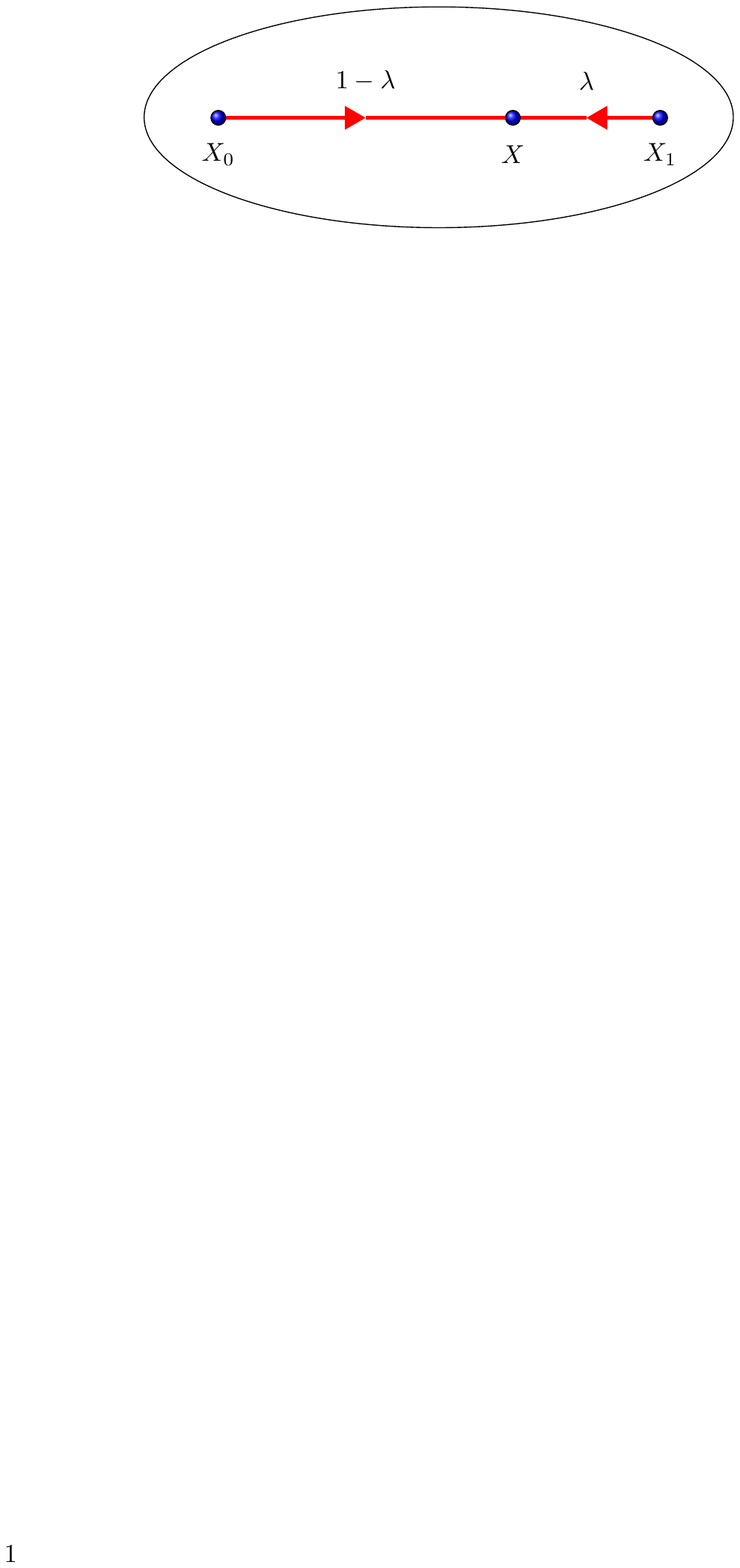}
\end{subfigure}%
\begin{subfigure}{.5\textwidth}
\hskip-8.5cm
  \centering
  \includegraphics[width=14cm]{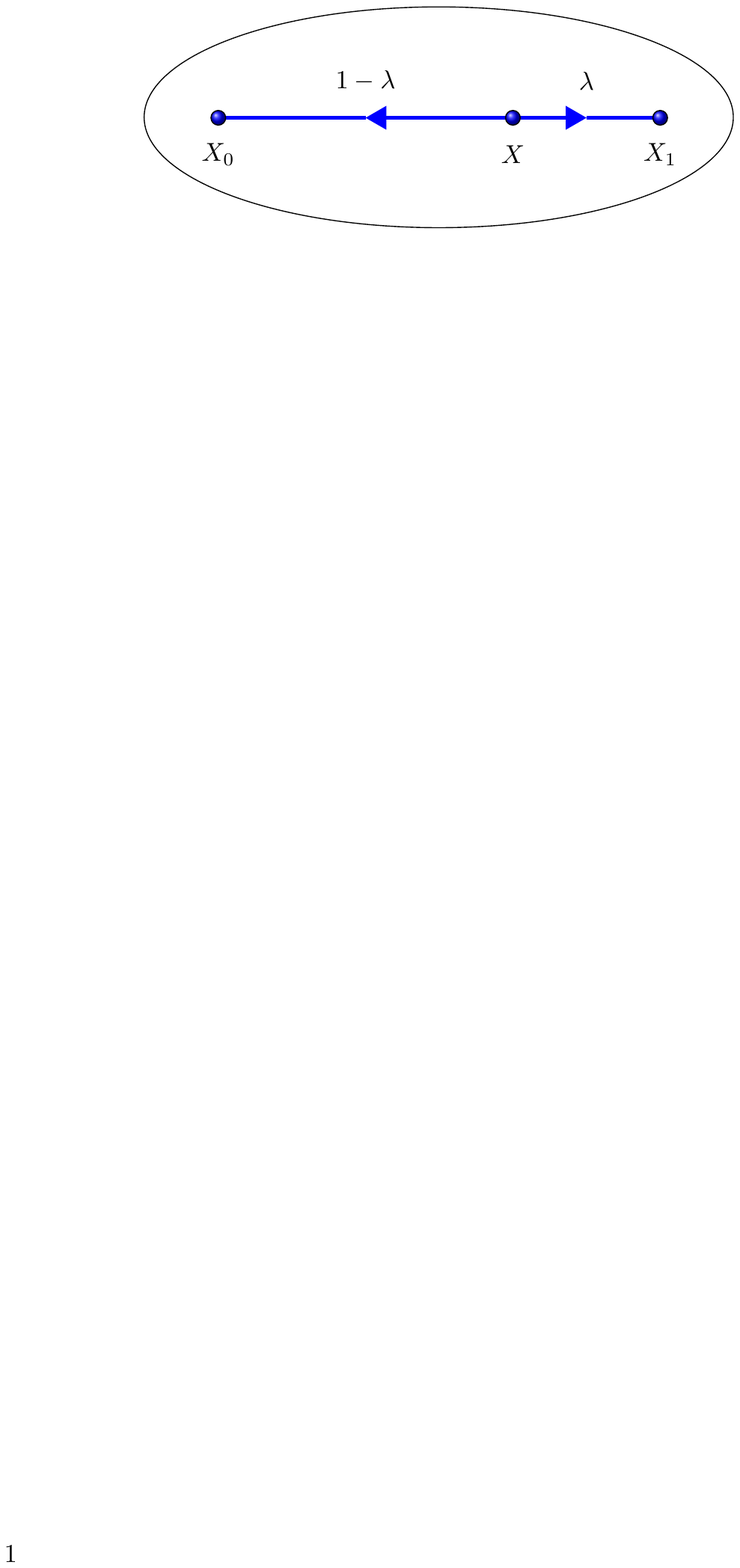}
\end{subfigure}\vskip-13cm
\caption{\small Definition of entropy, cf. Eqs. \eqref{1.17} and \eqref{1.18}. The left figure illustrates the processes employed for the definition by Eq. \eqref{1.17},  the right figure the analogous processes for \eqref{1.18}.}
\label{fig1}
\end{figure}


\begin{figure}[ht]
\centering
\includegraphics[width=5.5cm]{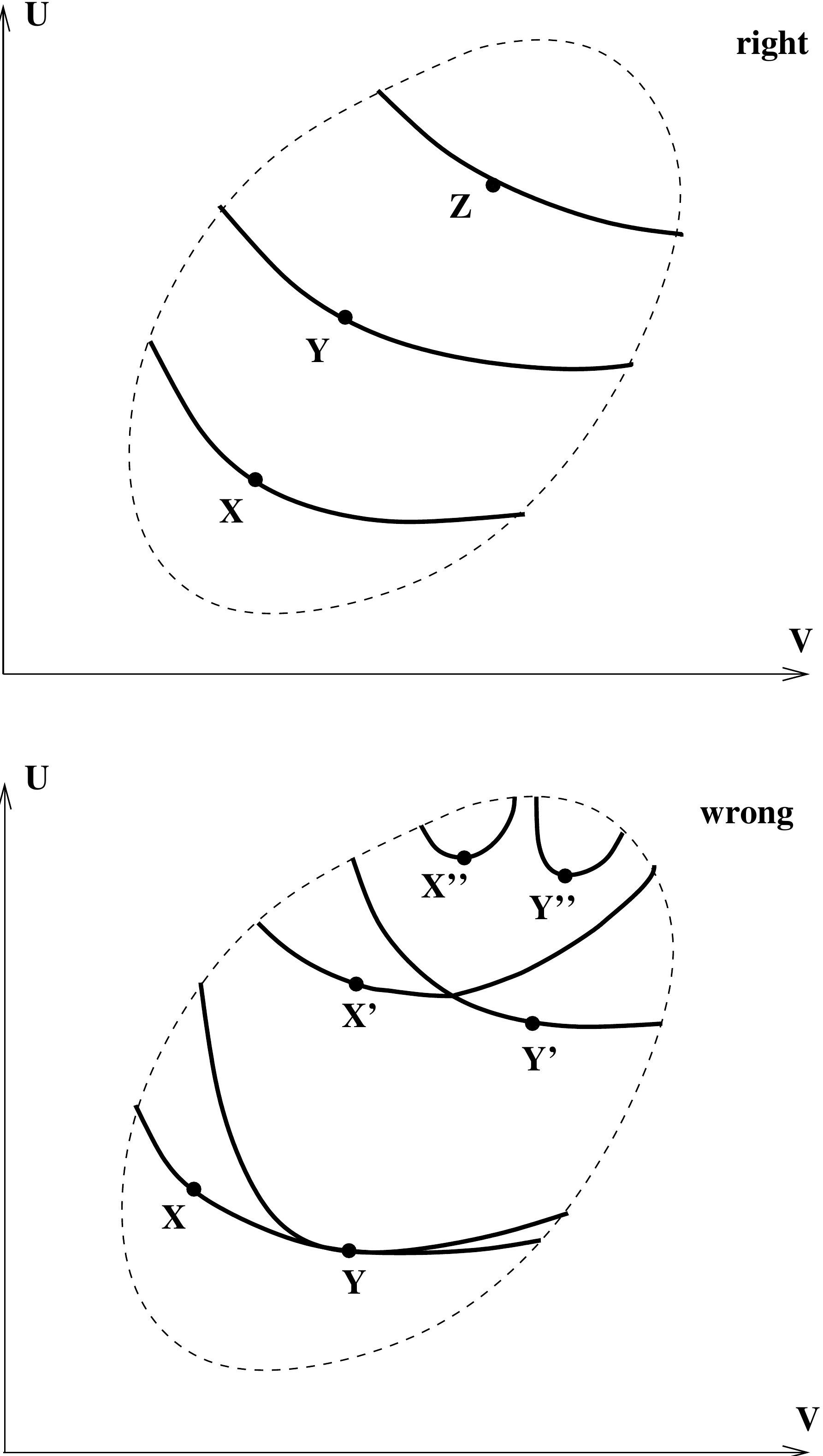}
\caption{The upper figure illustrates that forward sectors in the state space of a simple system (enclosed within the dashed curve) are nested.  The adiabat 
$\partial A_X$ is the hypersurface containing $X$, and the forward sector $A_X$ is the region above it. Likewise for the states $Y$ and $Z$. The lower figures shows what in principle could go wrong, but doesn't because of our axioms, in particular S2.\label{fig2}}
\end{figure}


\begin{figure}[ht]
\centering
\includegraphics[width=6cm]{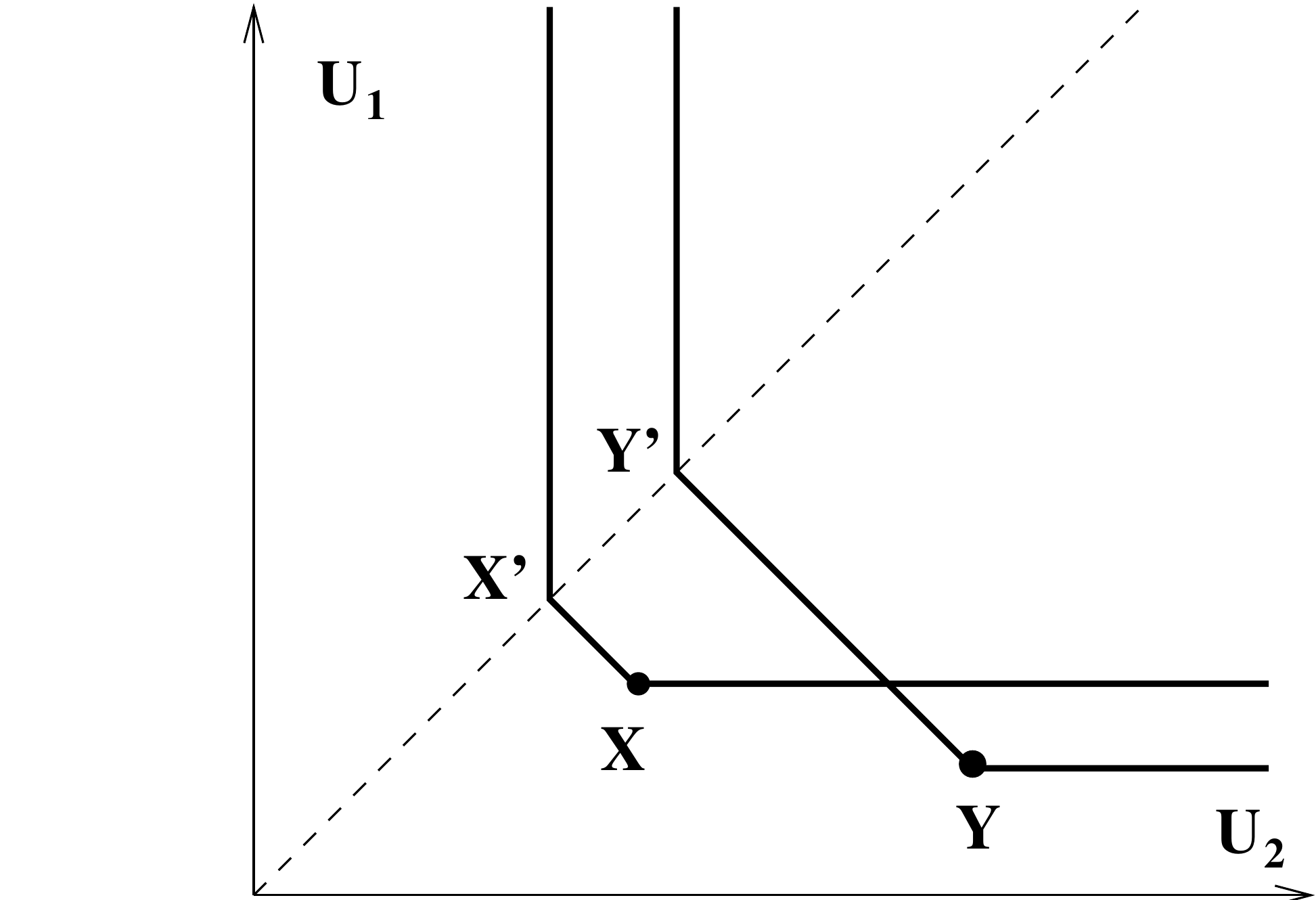}
\caption{The figure illustrates what can happen if the transversality axiom A4 does not hold. If the pressure is independent of $V$ isotherms and adiabats for a simple system $\Gamma$ coincide and the relation $\prec$ depends only on the energy. The figure shows  the state space of $\Gamma\times\Gamma$, ignoring the volumina, and two states, $X$ and $Y$, in $\Gamma\times\Gamma$. The thermal join $\Delta$ coreponds to the dashed line on the diagonal, $X'$ and $Y'$ are the projections the two states onto $\Delta$ as in \eqref{1.26}. It is evident from the figure that that $X$ and $Y$ are not comparable in $\Gamma\times\Gamma$.} {\label{fig3}}. 
\vskip1cm
\end{figure}

\begin{figure}[ht]
\centering
\includegraphics[width=6cm]{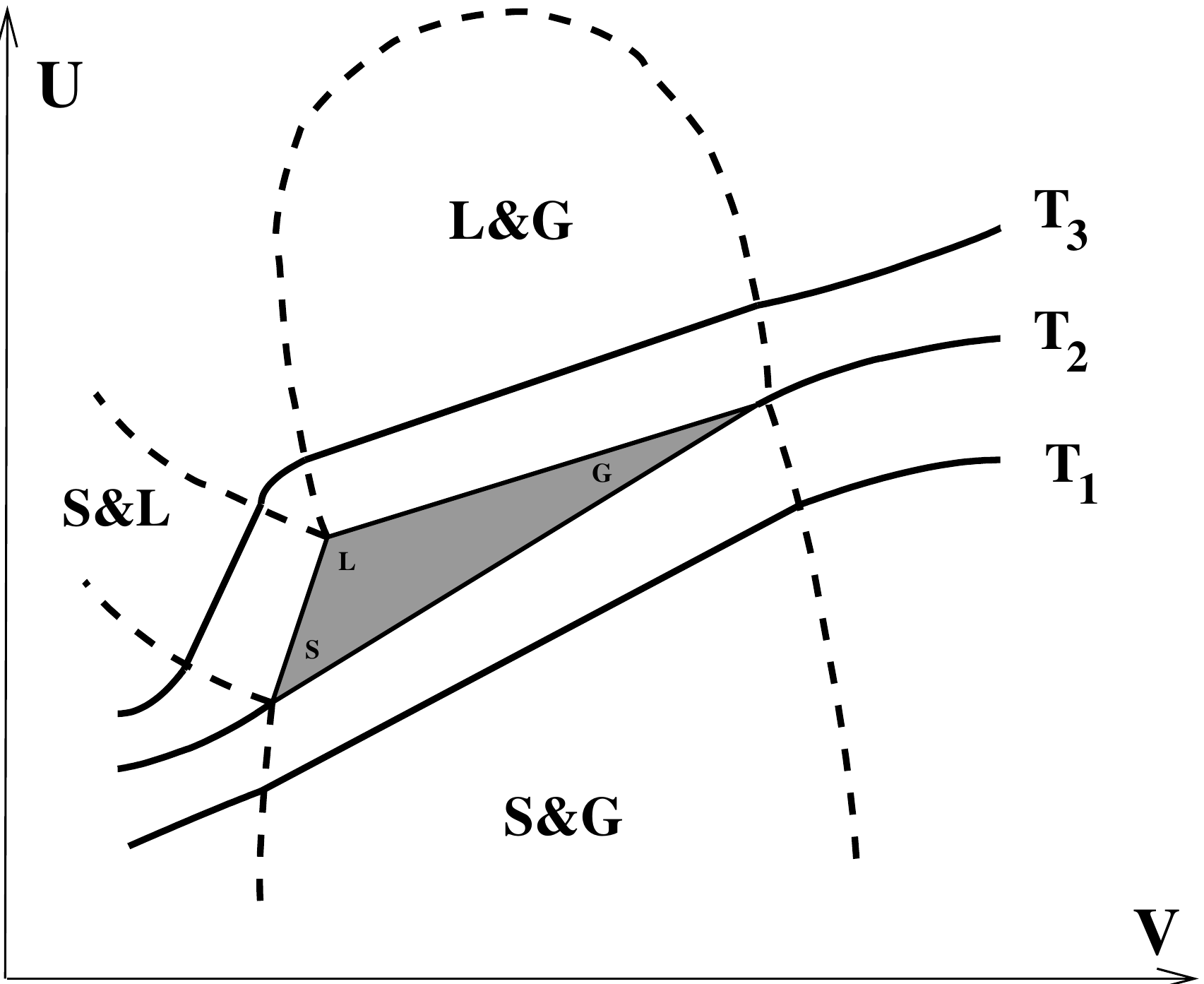}
\caption{The figure (not to scale) shows isotherms in the $(U,V)$ plane near the triple point of a system with three phases (solid S, liquid L, gas G). In the triple point region the temperature is constant which shows that isotherms need not have codimension 1.}{\label{fig4}}
\end{figure}
\fi

\end{document}